\documentclass[onecolumn]{emulateapj}
 \usepackage{graphicx}

\usepackage{amssymb}
\usepackage{amsbsy}
\usepackage{amsmath}
\usepackage{natbib}
\begin{document}
\title{Universality of electron distributions\\ in extensive air showers}
\email{andrzej.smialkowski@uni.lodz.pl}
 \author{Andrzej \'Smia\l{}kowski}
 \author{Maria Giller\altaffilmark{1}}
\affiliation{The University of Lodz, Faculty of Physics and Applied Informatics,\\ Pomorska 149/153, 90-236 Lodz, Poland}
\altaffiltext{1}{emeritus professor}

\begin{abstract}

Based on extensive air shower simulations it is shown that the electron distributions with respect to the two angles, determining electron direction at a given shower age, for a fixed electron energy and lateral distance, are universal. It means that the distributions do not depend on the primary particle energy or mass (thus, neither on the interaction model), shower zenith angle or shower to shower fluctuations, if they are taken at the same shower age. Together with previous work showing the universality of the distributions of the electron energy, of the lateral distance (integrated over angles) and of the angle (integrated over lateral distance) for fixed electron energy this paper completes a full universal description of the electron states at various shower ages. Analytical parametrizations of the full electron states are given. It is also shown that some distributions can be described by a smaller than five numbers of variables, the new ones being products of the old ones raised to some powers.\\
The accuracy of the present parametrization is sufficiently good for applying to showers with the primary energy uncertainty of 14$\%$ (as it is at the Pierre Auger Observatory). The shower fluctuations in the chosen bins of the multidimensional variable space are about 6$\%$, determining the minimum uncertainty needed for parametrization of the universal distributions. An analytical way of estimation of the effect of the geomagnetic field is given.\\
Thanks to the universality of the electron distributions in any shower a new method of shower reconstruction can be worked out from data of the observatories using the fluorescence technique. The light fluxes (both fluorescence and Cherenkov) for any shower age can be exactly predicted for a shower with any primary energy and shower maximum depth, so that the two quantities can be obtained by best fitting the predictions to the measurements
\end{abstract}
\section{Introduction}
Cosmic rays of the highest energies, $E_0 >10^{17}$ eV, can be detected and studied only by registering their effect produced in the atmosphere, i. e. the extensive air showers, cascades of particles (mainly electrons of both signs) and photons reaching the ground. Thanks to the fluorescence technique it is also possible to gain information about the state of a shower developing in the atmosphere at higher altitudes than the ground. Shower particles cause the atmosphere to emit fluorescence and Cherenkov light which intensity and spectrum depend on the electron state at the time of emission. Thus, when measured at many times while the shower traverses the atmosphere, the light characteristics enable a derivation of the total energy deposit of the shower particles changing with the atmospheric depth. This provides the best estimation of the shower primary energy i.e. the energy of the cosmic ray particle $E_0$. It also allows one to calibrate the much more frequent showers, registered by the ground detectors all the day round, not only at dark nights necessary for the light detectors measurement, as this was done for the first time at the Pierre Auger Observatory \citep{pao}.\\ 
However, an exact determination of the shower atmospheric profile, from which $E_0$ and $X_{max}$, the depth of the shower maximum, can be derived, demands a detailed knowledge of the state of shower most abundant particles, electrons, at any stage of its development.
Our present study shows that it is possible to obtain it, in contrast to the common belief that extensive air showers are a highly fluctuating phenomenon. Indeed, the atmospheric depths of the cosmic particle interactions, the energies transferred to the secondary particles, the consequent interactions of those particles are all subject to random processes. Thus, even if the primary particles are the same and have the same primary energies $E_0$ and the same zenith angles of the incidence to the atmosphere, the shower development through it will be different.\\
But it is the depth of the shower maximum, $X_{max}$, and the total number of electrons, $N(X)$, at various depths $X$ that will be different.
Continuing our and other author's work (see Section 2.1) we show that \emph{when referred to the shower maximum depth the function describing fully the distributions of all electron characteristics is the same in any shower, independently of the primary energy, mass, angle of incidence or fluctuations of shower development}.\\
By referring to the shower maximum we mean that instead of the depth $X$ one should use the age $s$ of the shower at this depth, defined as
\begin{equation}
s(X)=3X/(X+2X_{max})    \;\textnormal{.} 
\end{equation}
This was introduced for a description of shower development stage by Hillas \citep{hillas} by analogy to a pure electromagnetic cascade where this particular relation comes out of the analytical solutions.
It can be seen that the age $s$ depends actually on the ratio $X/X_{max}$, so that this very ratio could serve as a description of the shower development stage just as well. But we shall stay with the tradition and use formula (1) for it.\\
The state of an electron at some age $s$ can be described by four variables: the electron energy $E$, its distance to the shower axis in $\mathrm{g/cm}^2$ or in $r/r_M$, where $r_M$ is the Moli\`ere radius  at the depth $X(s)$, and the two angles of the electron direction with respect to the shower axis, the polar angle $\theta$ and the azimuth angle $\varphi$. 
Thus, our main conclusion reached in this paper for the first time, means that the five dimensional function $f(\theta,\varphi, r/r_M, E;s)$, representing the shape of the four-dimensional electron distribution at any level $s$, is the same for any shower, i.e. it has a universal dependence on the five variables (Section 2). \\
 This conclusion could be drawn for showers having enough electrons in small cells of the four-dimensional space of the variables $E, r/r_M, \varphi, \theta$ at various levels $s$, so that the function $f(\theta,\varphi, r/r_M, E;s)$ could be determined. This limits the shower primary energies to be greater than $(10^{16} - 10^{17} )$ eV, depending on the choice of the variable cell size and the range of the variable values to be well described.
We show (Section 3) that some of the electron distributions can be described by a number of variables smaller than five, the new variables being a product of the old ones raised to some power. This simplifies a lot the analytical parametrization of the electron distributions found here.  
In Section 4 we parametrize the simulated distributions by analytical functions and give an estimation of the accuracy of our parametrization. In Section 5 we describe an analytical way of estimation the distortions of the electron angular and lateral distributions caused by the geomagnetic field.\\
The universality of the electron distributions in showers provides a new method for reconstructing shower development in the atmosphere with the fluorescence technique (Section 6). The full description of the electron states and its universal character allows one to predict exactly the light fluxes (both fluorescence and Cherenkov) emitted by shower electrons at consequent time intervals (as recorded by telescopes), by adopting only two quantities for a shower: $N_{max}$ and $X_{max}$. Finding their values which fit best the data one can determine quite well the shower primary energy, and from the depth $X_{max}$ - obtain information about primary mass, e.g. \citep{abr1,abr2,abreu}. A discussion and summary constitutes the last Section 7.
\section{Universality of the angular distributions of electrons}
\subsection{Universality in earlier papers}
In a pure electromagnetic cascade the shape of the energy spectrum of electrons depends on the cascade age only, being independent of the primary energy; the result obtained analytically. By analogy, Hillas (1982) proposed that the same holds in hadronic extensive air showers, with his new definition of the shower age (Eq. 1). With the help of the shower simulation computer program CORSIKA (Heck et al. 1998) it was possible to show that electron energy spectra were the same in showers with different primary energies once the level considered corresponded to the same shower age (Giller et al. 2004, Nerling et al. 2006). The universal character of the energy spectra has been strenghtened by showing their independence of the primary particle (Giller at al. 2004), and thus, of the high energy interaction model used in the simulations.\\
Since the electron angles with respect to the shower axis depend on their energies (the main process responsible being the Coulomb scattering) the following study revealed the universality of the electron angular distributions (Giller et al. 2005a) i.e. their independence of the primary particle energy or mass. It was shown that integrated over electron energies and lateral distances the angular distributions depended on the shower age only. Next they were studied for fixed electron energies (Giller et al. 2005b) and it turned out that in this case the angular distributions did not depend even on the shower age. It should actually be expected since the electron angle depends mainly on the electron actual energy, and very weakly on its energy in the past.\\
A similar study was done later by Nerling et al.(2006) although their choice of rather large energy ranges was probably the reason for not quite confirming the independence of the angular distributions on shower age since the energy spectra within the ranges do depend on the age.\\
The idea for analysing electron distributions for their fixed energies was developed later by Lafebre et al. (2009) who studied various electron distributions in showers. They confirmed the independence of the angular distributions for fixed energies of the shower age and claimed their universal character with respect to the changes of the primary particle, its mass, energy or the incidence angle (although they expressed some doubt about they being universal against changes of the interaction model, what, in view of the independence of the primary mass, should not be doubtful). \\
The electron angular distribution does, of course, determine their lateral spread. It was shown (Giller et al. 2005a) that the lateral distribution of all electrons depended on the shower age only. Further
studies for fixed electron energies showed their universal character and a dependence on age only (Giller et al. 2007, Lafebre et al. 2009, Giller et al. 2015). Unlike the electron angle dependent on its actual energy the lateral distance does depend on the earlier electron history - even a small angle at much earlier level cause large lateral distance much below; thus earlier history is now important and the dependence on age follows. In Giller et al. (2015) a simple universal function of only one variable describing the lateral distribution at various ages for electrons with different energies was derived by rescaling the lateral distance.\\
 The universality of the electron distributions determined so far was used for prediction of the light fluxes emitted by showers, the detection of which is a method for studying high energy cosmic rays ($E_0 >10^{17}$ eV). The universal character of the lateral distribution of the energy deposit in the atmosphere, from which the lateral distribution of the fluorescence light emission can be straightforwardly obtained \citep{gora}, is a consequence of the universality of the lateral distributions of electrons with fixed energies. Nerling et al. (2006) and Giller $\&$ Wieczorek (2009) applied the universal angular distributions for deriving the production of the Cherenkov light in showers, although in an approximate way.\\
 The work described so far concerned distributions of one variable, electron angle or its lateral distance, integrated over the second one. However, it does not present an exact description of the full electron state in a shower since the two distributions are not independent. The electron angular distribution depends on their lateral distance. The first attempt to allow for this was made by Giller et al. (2015) where the dependence of the angular distribution on the lateral distance was considered. However, the distribution of polar angles was assumed to be independent of the azimuth (with respect to the shower axis) what was not quite true (see the present paper). Nevertheless, it was pointed out that thanks to the universal character of all electron distributions a new method for reconstruction of primary particle parameters from observation of fluorescence and Cherenkov light emitted by the shower can be worked out. To predict fluorescence emission it is enough to know electron lateral distributions for various energies; the angular distributions are irrelevant since this light is emitted isotropically. But to predict exactly the Cherenkov emission the angular distribution as a function of the electron lateral distance and energy must be known, particularly for showers observed from small distances. It is only in this paper that a description of the full electron state has been undertaken. 
\subsection{General considerations}
The state of electrons in a shower  is uniquely determined by  the numbers $\Delta N(\theta,\varphi, r/r_M, E,s)$ 
given for all possibly occupied regions of the five variables. Each $\Delta N$ is  the number of electrons with angles to the shower axis $(\theta, \theta +\Delta\theta)$, with azimuth angle $(\varphi, \varphi +\Delta\varphi)$ around the axis parallel to that of the shower, drawn at the position of the electron which is at a lateral distance to the axis ($ r/r_M, r/r_M+\Delta r/r_M,$), with energy ($ E,  E+\Delta E,$) and at a shower age $s$. The Moliere radius at a given height in the atmosphere is defined as
$r_M = (21\, \textrm{MeV}/\epsilon) l_0$, being the mean square scattering angle of electron with the critical energy $\epsilon$ $(\sim 82\, \textrm{MeV})$, along one radiation unit $l_0$.
We define a function $f(\theta,\varphi, r/r_M, E;s)$ as follows:
\begin{equation}
\Delta N(\theta,\varphi, r/r_M, E;s)=N(s)f(\theta,\varphi, r/r_M, E;s)\Delta \theta\Delta\varphi\Delta log\,(r/r_M) \Delta log E\;\textnormal{,} 
\end{equation}
where $N(s)$ is the total number of electrons at level $s$.
Note that the function $f(\theta,\varphi, r/r_M, E;s)$ has been defined for a single shower. However, we will show that for large showers, i.e. such that the numbers $\Delta N$ are large ($>100$), \emph{this function is universal}. This means that it is independent of the primary particle energy, its mass, the  incidence angle or even of the fluctuations in the shower development (thus the requirement  $\Delta N>100$). The function   $f$  can be represented as a product of functions, each  depending on one variable and some  parameters, as follows:
\begin{equation}
f(\theta,\varphi, r/r_M, E;s)=f_E(E;s)f_r(r/r_M;E,s)f_{\varphi}(\varphi;r/r_M,E,s)f_{\theta}(\theta;\varphi,r/r_M,E,s)     \;\textnormal{.} 
\end{equation}
The variables on the r.h.s. of the semicolons are the parameters of the functional dependence on the variables before the semicolons. The functions $f_E,f_r,f_{\varphi}$ and  $f_{\theta}$ are normalized to unity when integrated over $log\,E$, $log\,(r/r_M)$, $\theta$ and $\varphi$ respectively. They are correspondingly the distributions : of electron energy $E$ at a given $s$, of distance $r/r_M$  at given $E$ and $s$, of the angle $\varphi$ at given $r/r_M$, $E$ and $s$, and of $\theta$ at given $\varphi$, $r/r_M$, $E$, and $s$.
As it has been described above, the universality of the electron energy spectra and lateral distributions for various energies was already proven. It is seen
 from Eq. 3 that  to prove  the universality of \emph{any} electron distribution, represented by $f(\theta,\varphi, r/r_M, E;s)$, it is left to demonstrate that  each of the functions $f_{\varphi}(\varphi;r/r_M,E,s)$ and $f_{\theta}(\theta;\varphi,r/r_M,E,s)$ 
are universal.\\
To obtain the electron distributions we simulated showers with CORSIKA \citep{cors}, version 7.4 with QGSJET-II model for high-
energy interactions. As an atmospheric model was used The US Standard Atmosphere.
However, as it will be claimed, our conclusions do not depend on the above choices.\\
CORSIKA is a much elaborated computer code for simulating the development of the air showers in the atmosphere. Choosing the primary particle, its energy and the incidence angle in the atmosphere it follows all the produced secondary particles (hadrons, muons, electrons, photons and neutrinos) and their interactions. Of course, for hadrons a high energy interaction model has to be adopted. The most abundant particles in a shower are electrons (both signs) and it is their state i. e. distributions over several variables at various atmospheric levels that we are concerned about in this work.\\
We find the number of electrons $\Delta N(\theta, \varphi,r/r_M,E;s)$ in variable bins
$\Delta\theta \Delta\varphi\Delta (log\,r/r_M)\Delta log\,E$ at various ages $s$.  Our study is restricted to variable regions where there are most electrons in the shower, i.e. to  $0.7< s <1.3$,  $20\,\textrm{MeV}< E <200\, \textrm{MeV}$ and $r/r_M<1.5$.  The lower energy limit was chosen mainly by our final interest to describe the shower Cherenkov radiation with the threshold $E=21$ MeV at sea level and increasing with height. For $E<20$ MeV the accurate electron angular distributions are not necessary for the purpose of reconstructing shower properties from the optical data because the fluorescence light is emitted isotropically. However, our parametrizations fit well down to $\sim 15$ MeV.\\ 
Single showers with $E_0 =10^{17}$ eV and $10^{16}$ eV were fully simulated i.e. without using the thinning procedure \citep{hillas2}, whereas at $10^{18}$ and $10^{19}$ eV  the thinning (a procedure constraining the number of followed particles by assigning to some of them a weight) was, of course, used.\\
\begin{figure}[!htb]
        \centering
        \includegraphics[width=4.5in]{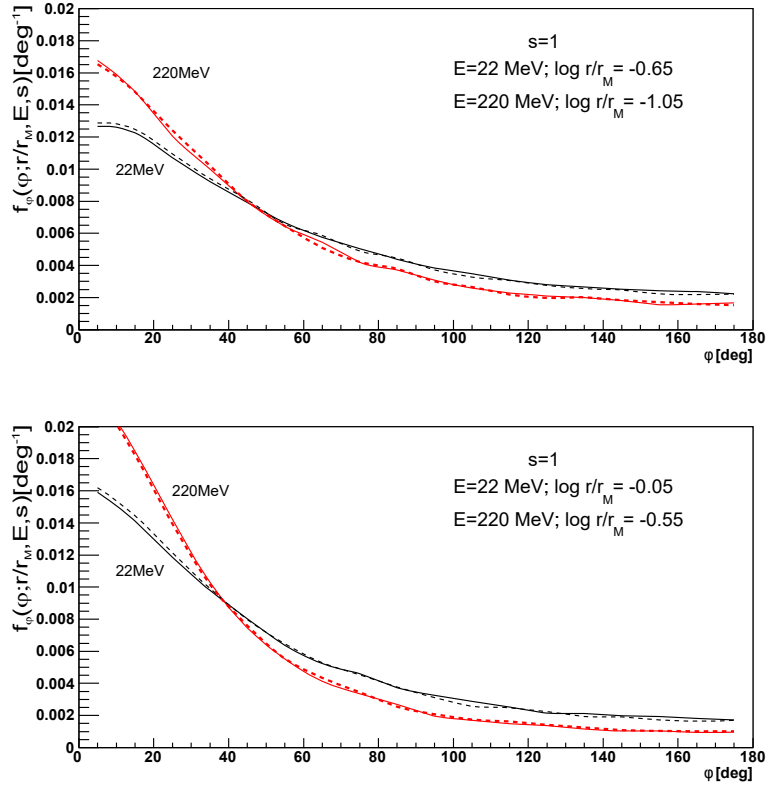}
 \caption{Distributions $f_{\varphi}(\varphi;r/r_M,E,s)$ of electron azimuth angle $\varphi$, integrated over polar angles $\theta$ with respect to the shower axis, for two electron energies $E$ within $\Delta \, logE= 0.15$, at two lateral distances $r/r_M$ within $\Delta log (r/r_M)=0.1$, obtained from shower simulations with CORSIKA. Continuous curves - one primary proton shower with $E_0=10^{19}$ eV, dotted curves - average of 10 iron showers with $E_0=10^{17}$ eV; $s=1$. Independence of the primary particle's energy or mass is seen.  $\varphi=0$ means direction away from shower axis.}
 \label{fig1}
 \end{figure}

\begin{figure}[!htb]
    \begin{minipage}{0.5\textwidth}
        \centering
        \includegraphics[width=3.6in]{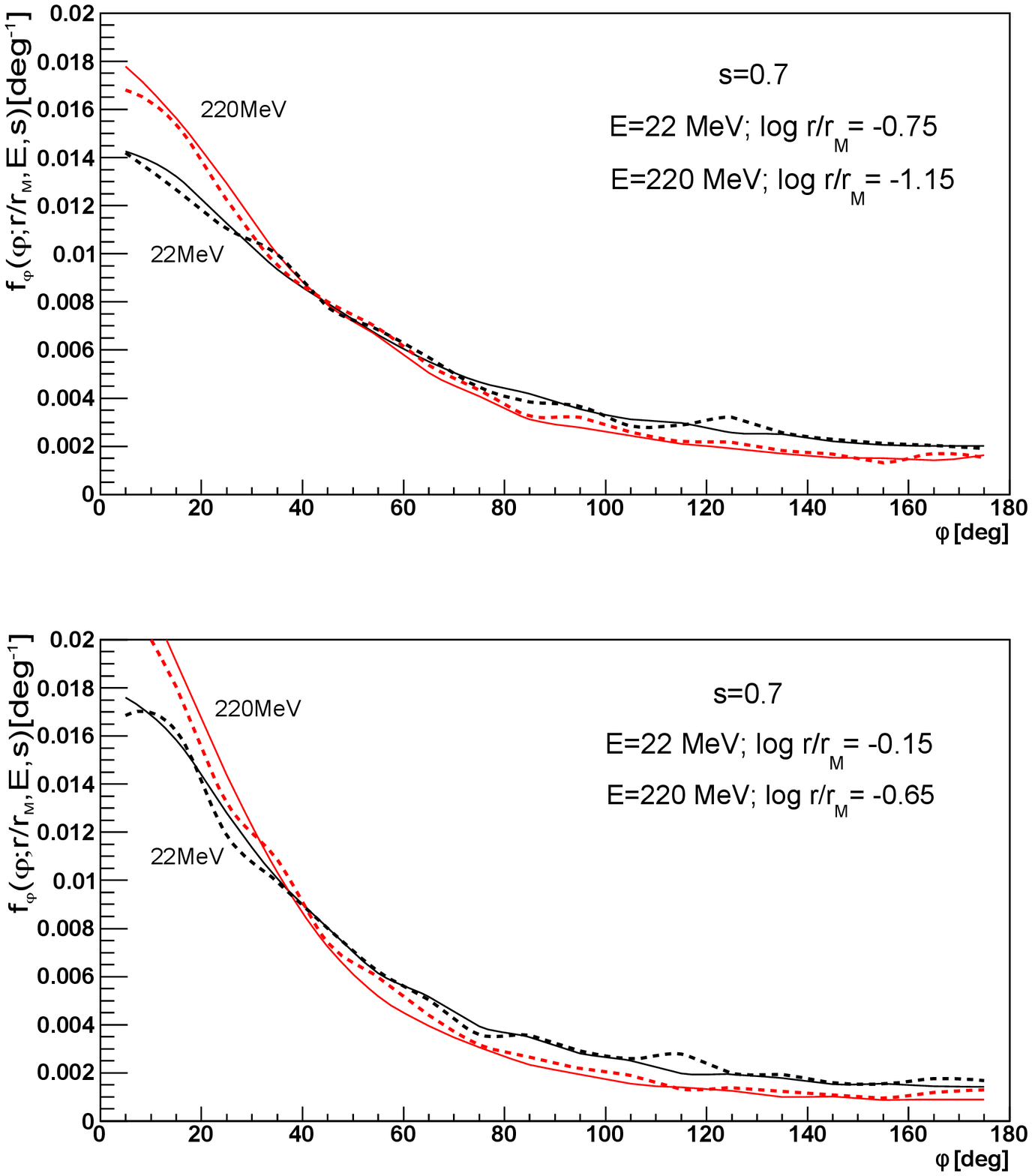}

    \end{minipage}
    \begin{minipage}{0.5\textwidth}
        \centering
        \includegraphics[width=3.6in]{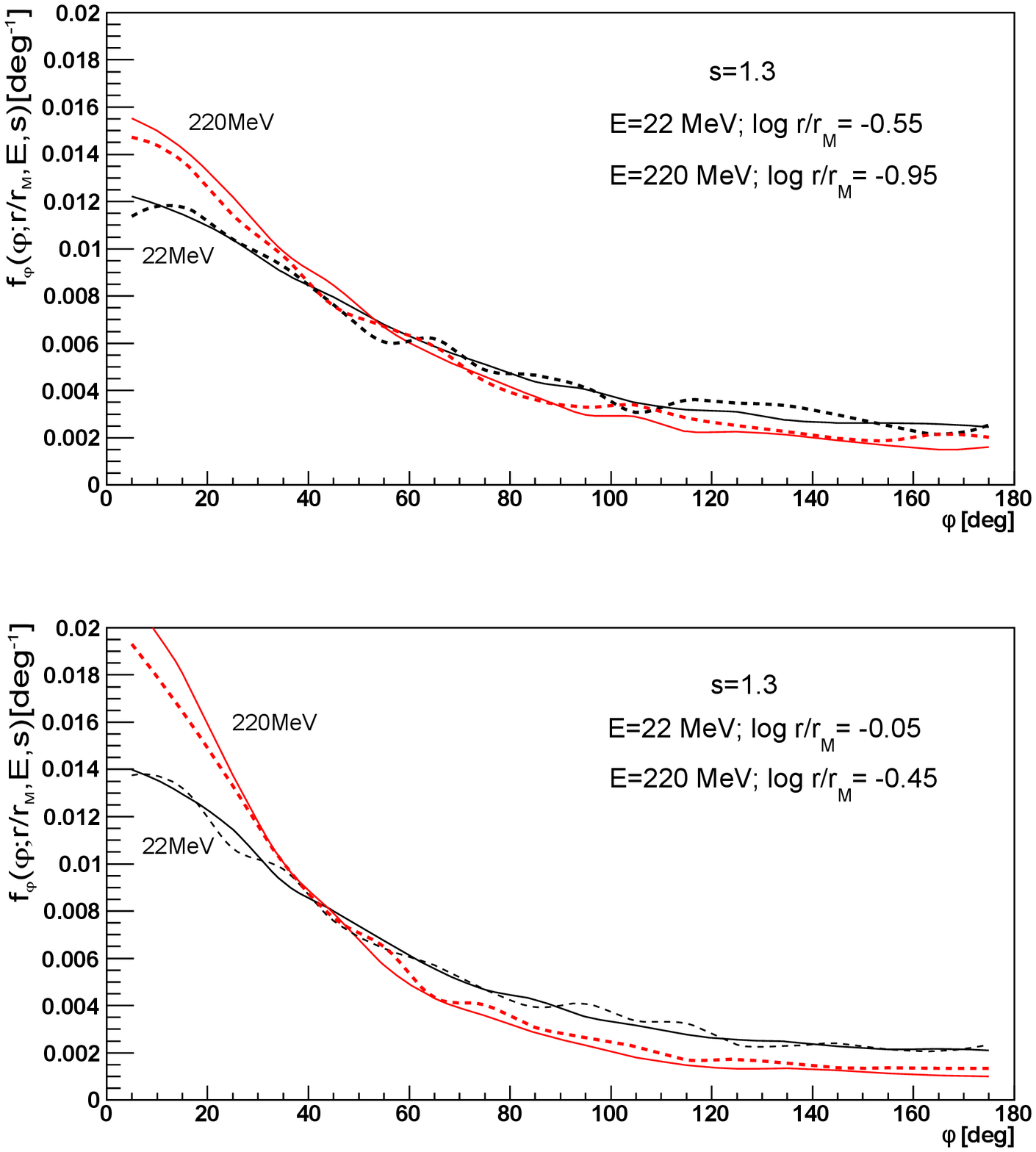}

    \end{minipage}
  \caption{Distributions $f_{\varphi}(\varphi;r/r_M,E,s)$ as in Fig. 1 but away from shower maximum. Left column: $s=0.7$; right: $s=1.3$.}
 \label{fig2}
\end{figure}
\subsection{Distributions of azimuth angles}
Figs 1 and 2 demonstrate the universality of the distributions
 $f_{\varphi}(\varphi;r/r_M,E,s)$ of the azimuth angle $\varphi$ of electrons with two fixed energies $E = 22$ MeV and $220$ MeV, at  two distances $r/r_M$  for each $E$,  at three shower ages $s = 1,\, 0.7$ and $1.3$.
 The  distributions refer to \emph{one} proton shower with $E_0=10^{19}$ eV and to the average of 10 iron showers with $E_0=10^{17}$ eV (actually the fluctuations from shower to shower  for $E_0=10^{17}$ eV were not large, but for a better comparison with the proton shower we have averaged them over 10 showers). We see that essentially there is no difference between the two curves in any of the graphs of Fig. 1 and 2.
The particular values of $E$ and $log (r/r_M)$ have been chosen in such a way as to correspond  to two values on both sides of the maximum of the $log E$, and correspondingly $log(r/r_M)$, distributions (the latter  describes fraction of electrons in \emph{rings} with a constant thickness $\Delta log(r/r_M)=0.1$). At maximae of these distributions the independence of the primary particle characteristics  is even better.
\subsection{Distributions of  polar angles}
Next, we go to the function, $f_{\theta}(\theta; \varphi,r/r_M,E,s)$, to be checked on the universality - the distribution of angles $\theta$ for fixed values of the parameters $\varphi,r/r_M,E,s$. Figs 3 and 4 illustrate the independence of the polar angle $\theta$ distributions of electrons $f_{\theta}(\theta; \varphi,r/r_M,E,s)$ of the primary particle energy and mass for $s=1$.  The  chosen regions of the azimuth angle correspond to the electrons deflected mostly away   from  the shower axis: $0<\varphi<20^{\circ}$,  toward it: $160^{\circ}<\varphi< 180^{\circ}$, and to those deflected perpendicularly to $\vec{r}$ : $70^{\circ}<\varphi<90^{\circ}$ and $\,90^{\circ}<\varphi<110^{\circ}$.\\
\begin{figure}[!htb]
\centering
        \includegraphics[width=4.8in]{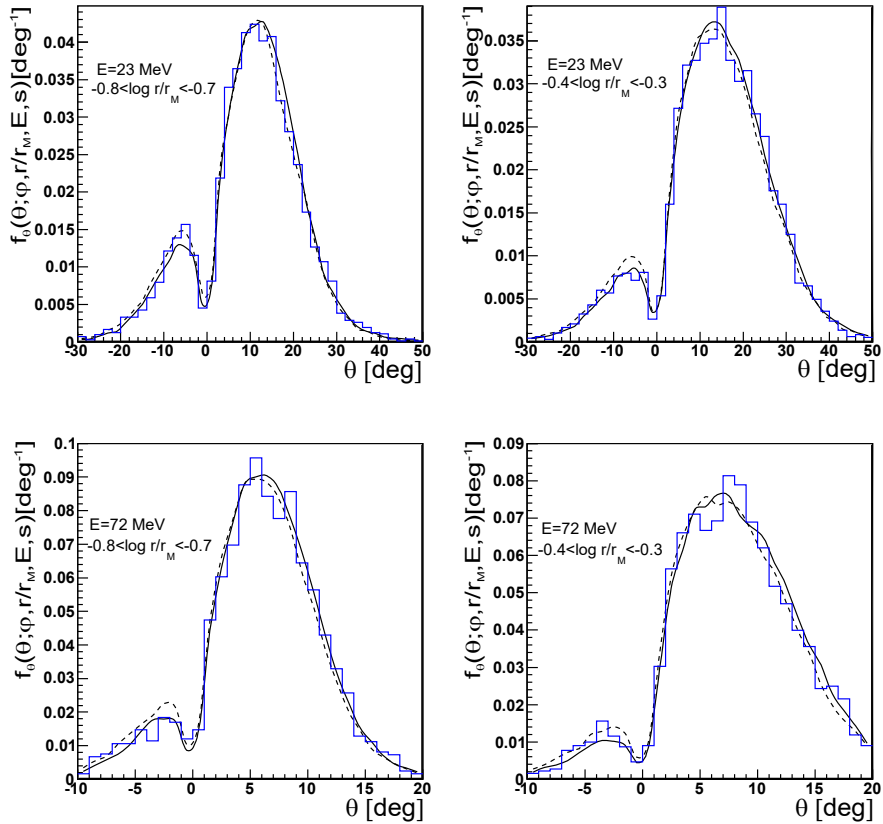}
  \caption{
Distributions $f_{\theta}(\theta; \varphi,r/r_M,E,s)$ of electron angle $\theta$ (in deg) for $s=1$ for two electron energies $E$ and two distances $r/r_M$, obtained from shower simulations with CORSIKA.
Continuous curve - one primary proton shower with $10^{19}$ eV, dashed - average of 10  Fe $10^{16}$ eV showers, histogram - one Fe $10^{16}$ eV.
 $0<\varphi<20^{\circ}$ - bigger peaks, $160^{\circ}<\varphi<180^{\circ}$ - smaller peaks.}
 \label{fig3}

\end{figure}
\begin{figure}[!htb]
\centering
        \includegraphics[width=4.8in]{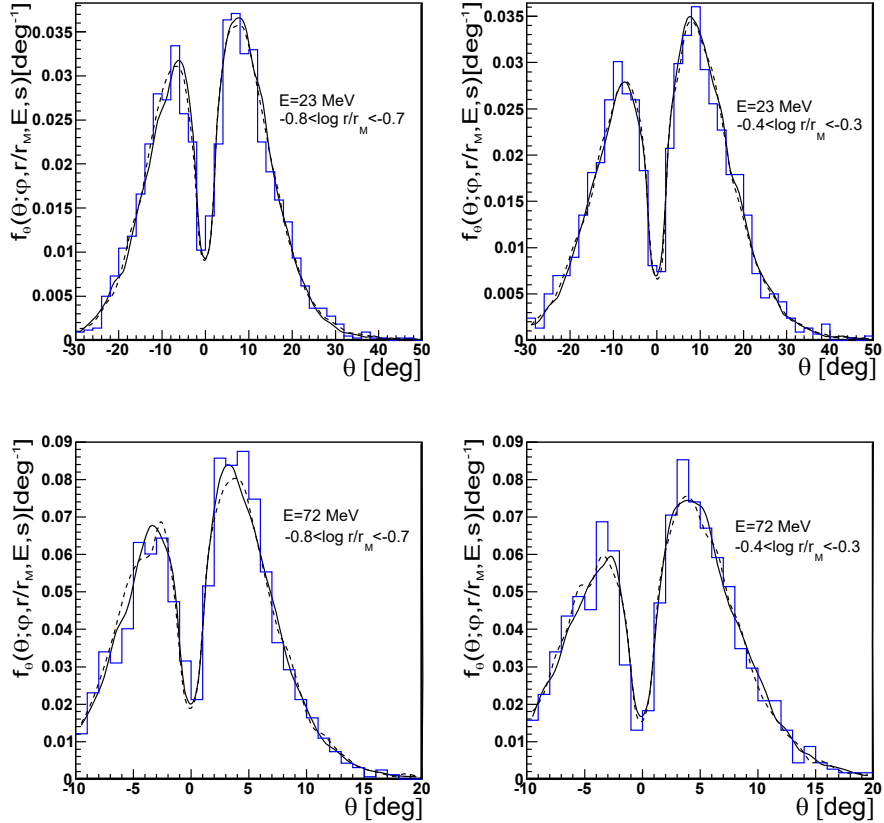}

  \caption{Distributions $f_{\theta}(\theta; \varphi,r/r_M,E,s)$  as in Fig. 3, but for angular directions perpendicular to $\vec r$. $70^{\circ}<\varphi<90^{\circ}$ - bigger peaks, $90^{\circ}<\varphi<110^{\circ}$ - smaller peaks.}
 \label{fig4}
\end{figure}

\begin{figure}[!htb]
        \centering
        \includegraphics[width=4.8in]{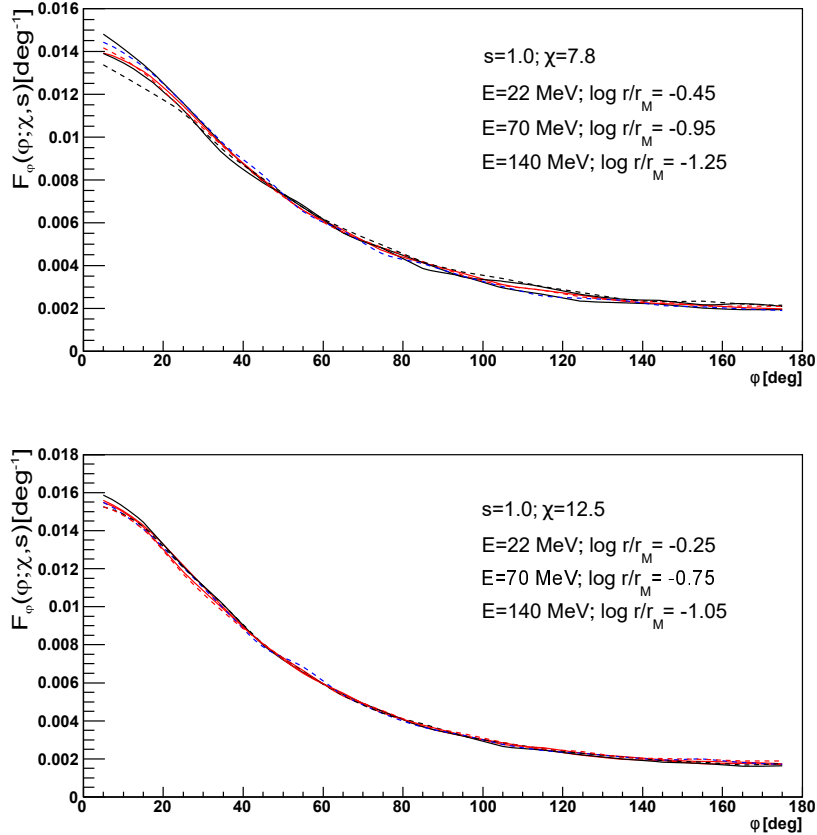}
  \caption{Distributions $F_{\varphi}(\varphi;\chi,s)$ of the electron azimuth angle $\varphi$ for three electron energies $E$, at lateral distances $r/r_M$ that correspond to  $\chi=7.8$ (upper graph) and  $\chi=12.5$ (lower graph). Continuous curves - one proton shower with $E_0=10^{17}$ eV, dotted - iron shower with $E_0=10^{17}$ eV. $s=1$. Independence of $E$ and $r/r_M$ for constant $\chi$ is seen, as well as the universality with respect to the primary.}
 \label{fig5}
\end{figure}

\begin{figure}[!htb]
        \centering
        \includegraphics[width=4.8in]{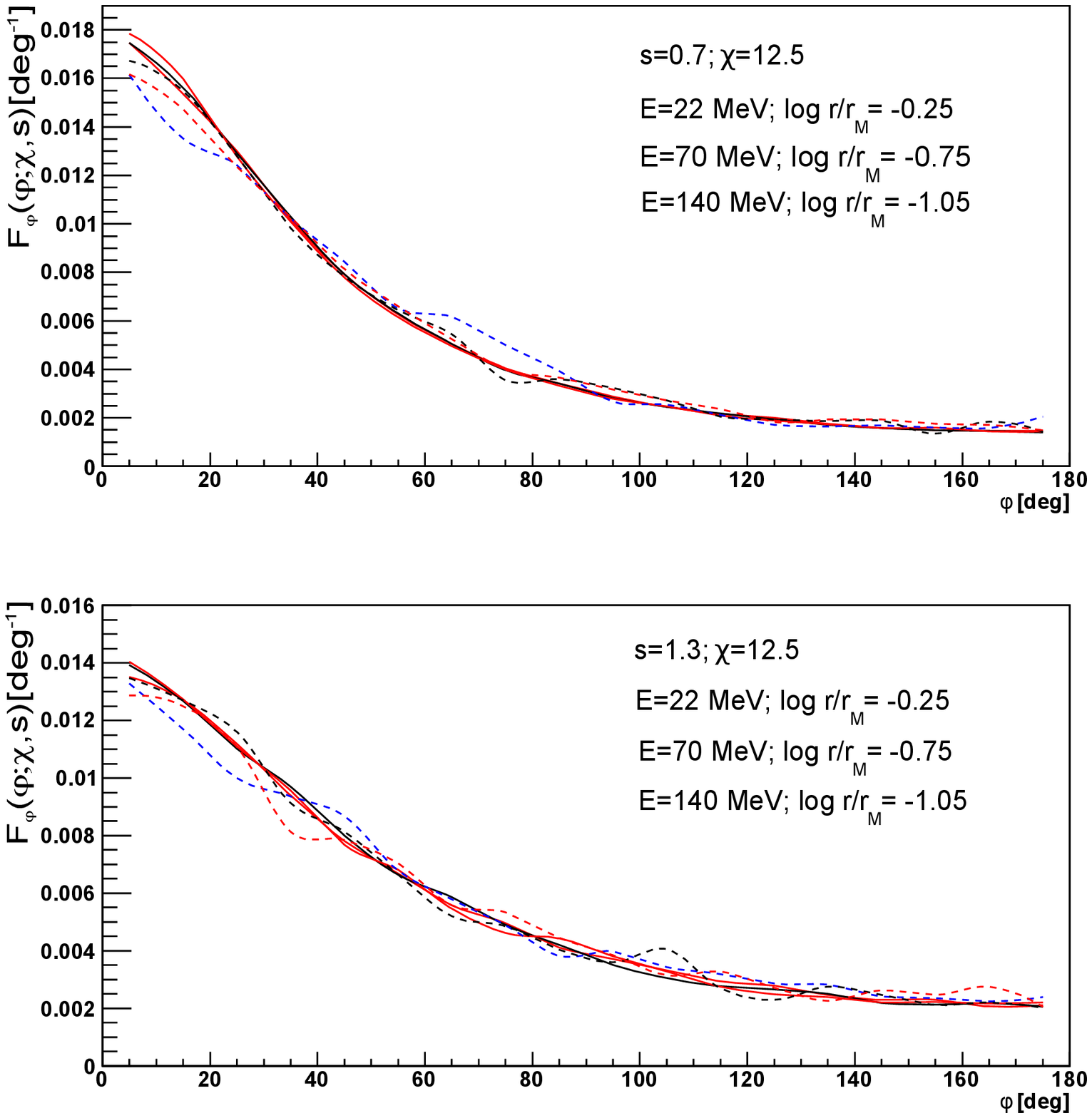}  

  \caption{As in Fig. 5 but for $\chi=12.5$. Continuous curves - one primary proton shower with $E_0=10^{19}$ eV, dotted - average of 10 iron showers with $E_0=10^{17}$ eV. Upper graph: $s=0.7$, lower graph: $s=1.3$.}
 \label{fig6}
\end{figure}

The average curve from 10 iron showers with the primary energy $E_0 =10^{16}$ eV agrees very well with that referring to a single proton shower with  $10^{19}$ eV.  Comparing the average distribution with that for a single shower one can also see fluctuations in individual bins which are partly due to small electron numbers in the bins (see Section 4.3). 
We have verified that such universality of the distributions $f_{\theta} (\theta; \varphi , r / r_M , E , s )$ holds at every stage of the shower development in the range of $0.7< s <1.3$ .
These angular distributions,  when taken for the same values of energy and lateral distances of electrons (in units of Moli\`ere radius)  do show dependence on the shower development stage.\\

\section{Extended (internal) universality of the angular distributions}
\begin{figure}[!htb]
   
    \begin{minipage}{0.48\textwidth}
        \centering
        \includegraphics[width=3.8in]{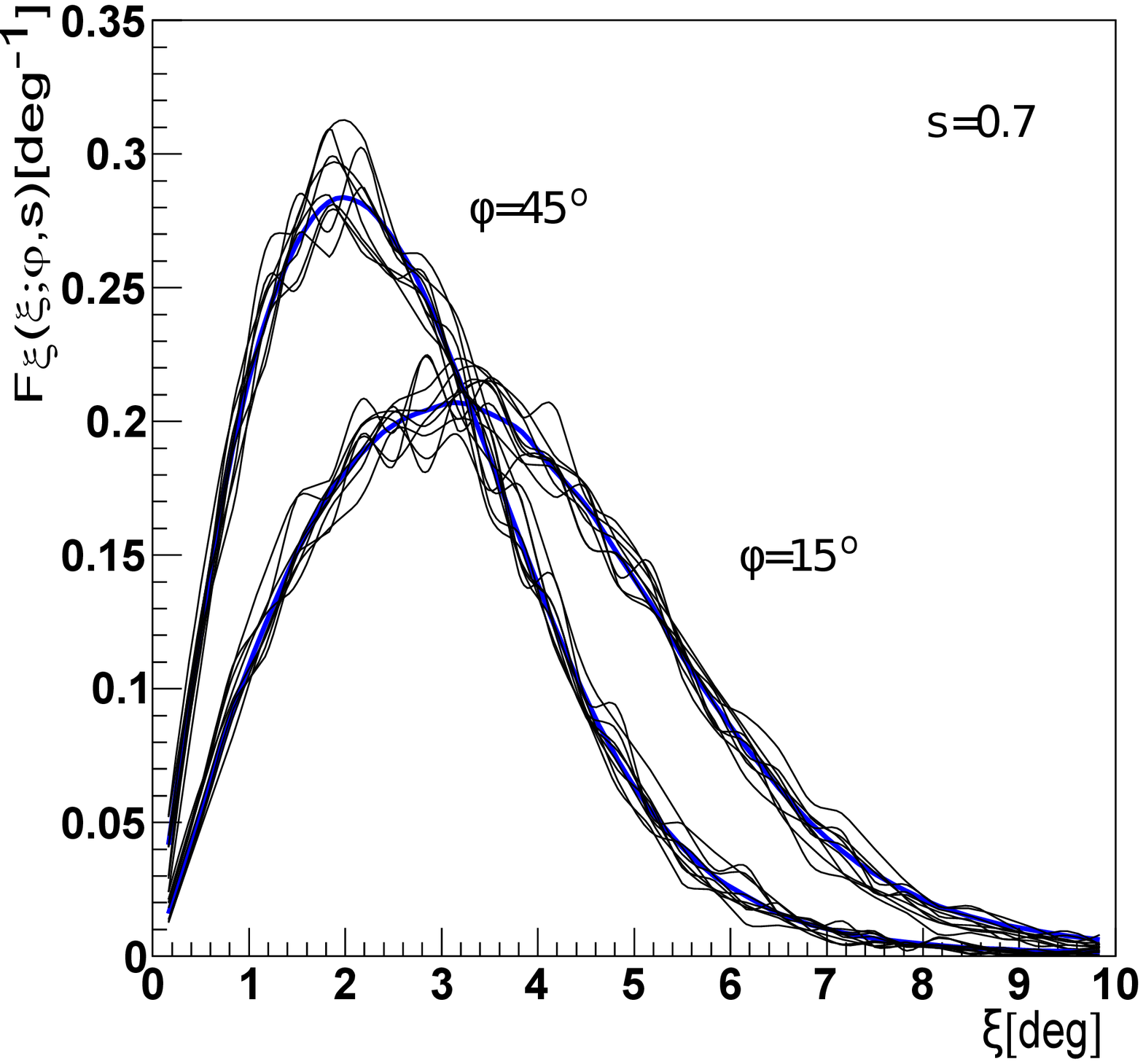}

    \end{minipage}%
    \begin{minipage}{0.48\textwidth}
        \centering
        \includegraphics[width=3.8in]{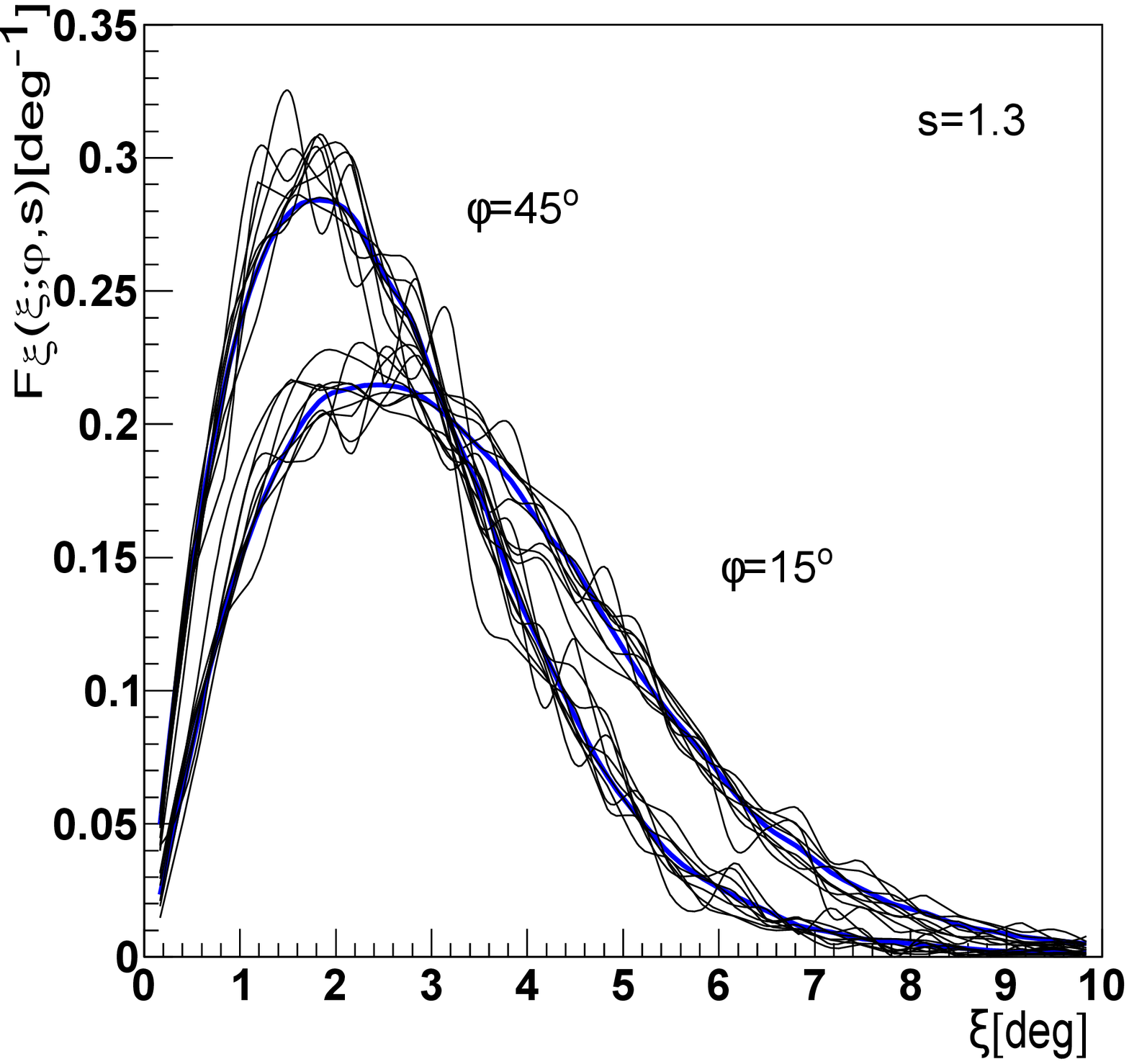}

    \end{minipage}
  \caption{Distributions of $F_{\xi} (\xi ; \varphi , s )$ for $s=0.7$ and $s=1.3$. Nine curves in each group refer to combinations of three energies and three lateral distances for each region of azimuth angle $\varphi$ (see text). Thick single curves correspond to distributions for all energies, distances and $\theta$ angles. One $10^{19}$ eV proton shower.}
 \label{fig7}
\end{figure}
Studying the electron angular distributions in multidimensional space of polar and azimuth angles, lateral distances, energies and ages of showers, allows us also searching  for a possible internal type of universality.
This means that the angular distributions may depend on a combination of some of the five independent variables rather than on each of the variable separately.
 Such a universality would simplify the description of the angular distributions by diminishing the number of necessary variables. A similar successful search was done earlier while describing the lateral distributions of electrons with various energies independent of their angles \citep{gsw}\\

\subsection{Variable $\chi$}
We have found that introducing a new variable
\begin{equation}
  \chi=(E/1\,\textrm{MeV})\cdot r/r_M  \;\textnormal{,}
\end{equation} 
the distributions $F_{\varphi} (\varphi ; \chi, s )$, where $F_{\varphi} (\varphi ; \chi, s )d\,\varphi$ gives the fraction of the number of electrons with angles $(\varphi, \varphi +d\,\varphi)$ at level $s$, with $E\cdot r/r_M=\chi$, with respect to the total number of them with any $\varphi$,  have almost universal shapes. Thus, the function $f_{\varphi} (\varphi ; r / r_M , E , s )$ reduces to a function $F_{\varphi} (\varphi ; \chi, s )$  of three variables: $\chi , \varphi$ and $s$. We have chosen $E$ in MeV, so that $\chi$ is a dimensionless variable.  \\
Fig. 5  represents the universality of the distributions  $F_{\varphi}(\varphi;\chi, s=1)$ of the azimuth angle $\varphi$ of electrons with energies $E=22$ MeV, 70 MeV and 140 MeV and corresponding lateral distances for two values of $\chi=7.8$ and $12.5$. 
Curves are for one proton and one iron shower with energy $10^{17}$ eV without thinning.  
In  Fig. 6 we show how the universality holds in the case of a young ($s=0.7$) and an old ($s=1.3$) shower, this time taking electron distributions for one $10^{19}$ eV proton shower  and ten $10^{17}$ eV iron showers  with thinning procedure switched on. The energies and lateral distances refer to $\chi=12.5$. 
Fig. 5 and 6 also demonstrate the universal behaviour of  $F_{\varphi} (\varphi ; \chi, s )$ with respect to the change of the primary particle energy or mass.  
\subsection{Variable $\xi$}
It is not only that the distributions $F_{\varphi} (\varphi ; \chi, s )$ are universal but also the distributions of the polar angle $\theta$ show a similar character.
We have searched for  \emph{one} variable which could combine the energy, lateral distance and $\theta$ angle for constant $\varphi$ and $s$. 
We have found a variable $\xi$ defined as
\begin{equation}
\xi=\frac{(E/1\,\textrm{GeV})^{\alpha} (1+r/r_M)}{(r/r_M)^{\beta}}\cdot \theta  \ \textnormal{,}
\end{equation}
where $\alpha$ and $\beta>0$.  We choose $E$ in GeV and $\theta$ in degrees.
Using this variable we get for all regions of the electron energies and lateral distances almost identical shapes of the normalized distributions of  $\xi$ for a given azimuth angle $\varphi$ and shower age $s$.  
Thus, the function $f_{\theta} (\theta ; \varphi , r / r_M , E , s )$ of five variables reduces to a function $F_{\xi} (\xi ; \varphi , s )$ of only three variables: $\xi , \varphi$ and $s$, where $F_{\xi} (\xi ; \varphi , s )d\,\xi$ is the fraction of the number of electrons with parameter $(\xi, \xi +d\,\xi)$, the angle $\varphi$ at level $s$, with respect to the total number of them with any $\xi$.  \\
Using a procedure of minimizing differences between histograms binned for various energies and distances in different $\varphi$ ranges we  have found the best values of $\alpha$ and $\beta$ (see Appendix Eq. A6).\\
Fig. 7 shows the distributions
  $F_{\xi} (\xi ; \varphi , s )$ for two values of $\varphi$ , at $s = 0.7$ and $1.3$ (at $s=1$ the picture is similar with less fluctuations). For each $\varphi$ there are nine curves,
referring to combinations of three electron energies: $E=23$, $70$ and $200$ MeV and three lateral distances: $log ( r / r_M )=-1.25$, $-0.85$ and $-0.45$ for $s=0.7$ and $log ( r / r_M )=-1.05$, $-0.65$ and $-0.25$ for $s=1.3$.
The different sets of the lateral distances reflect the fact that in older showers the lateral distribution becomes flatter.
 It is seen that the distributions are essentially independent of energy, lateral distance or the polar angle once the value of $\xi$ is fixed.\\

\section{Analytical parametrization of the universal angular distributions}

\begin{figure}[!htb]
    \centering

        \includegraphics[width=4.7in]{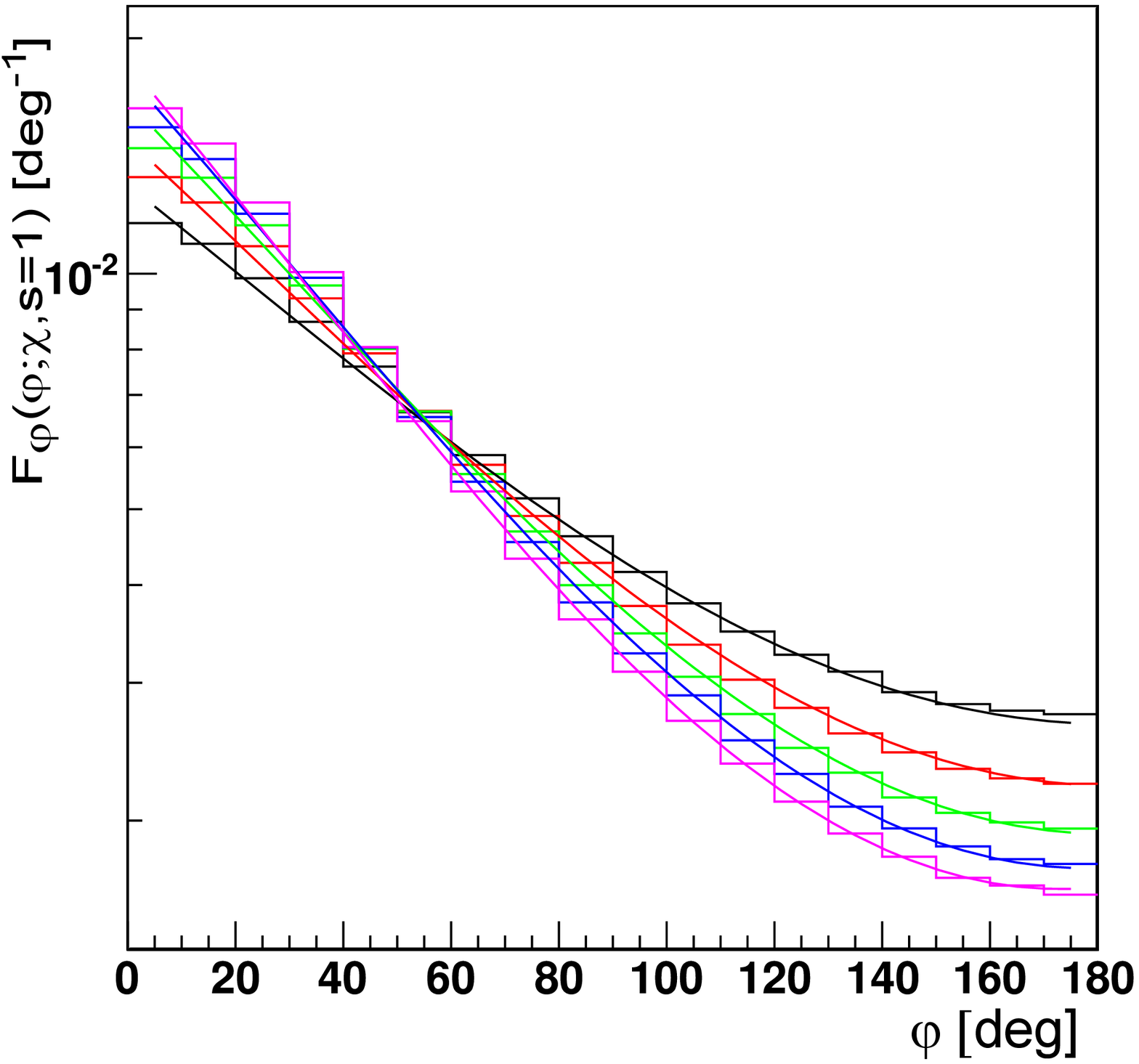}

  \caption{Comparison of simulated (histograms) and parametrized (lines) distributions $F_{\varphi}(\varphi;\chi,s=1)$ of azimuth angles $\varphi$ for five equidistant values of $\chi$ from 3 to 15 (top to bottom at the right side). One $10^{17}$ eV iron shower. }
 \label{fig8}
\end{figure}

\begin{figure}[!htb]
        \centering
        \includegraphics[width=4.8in]{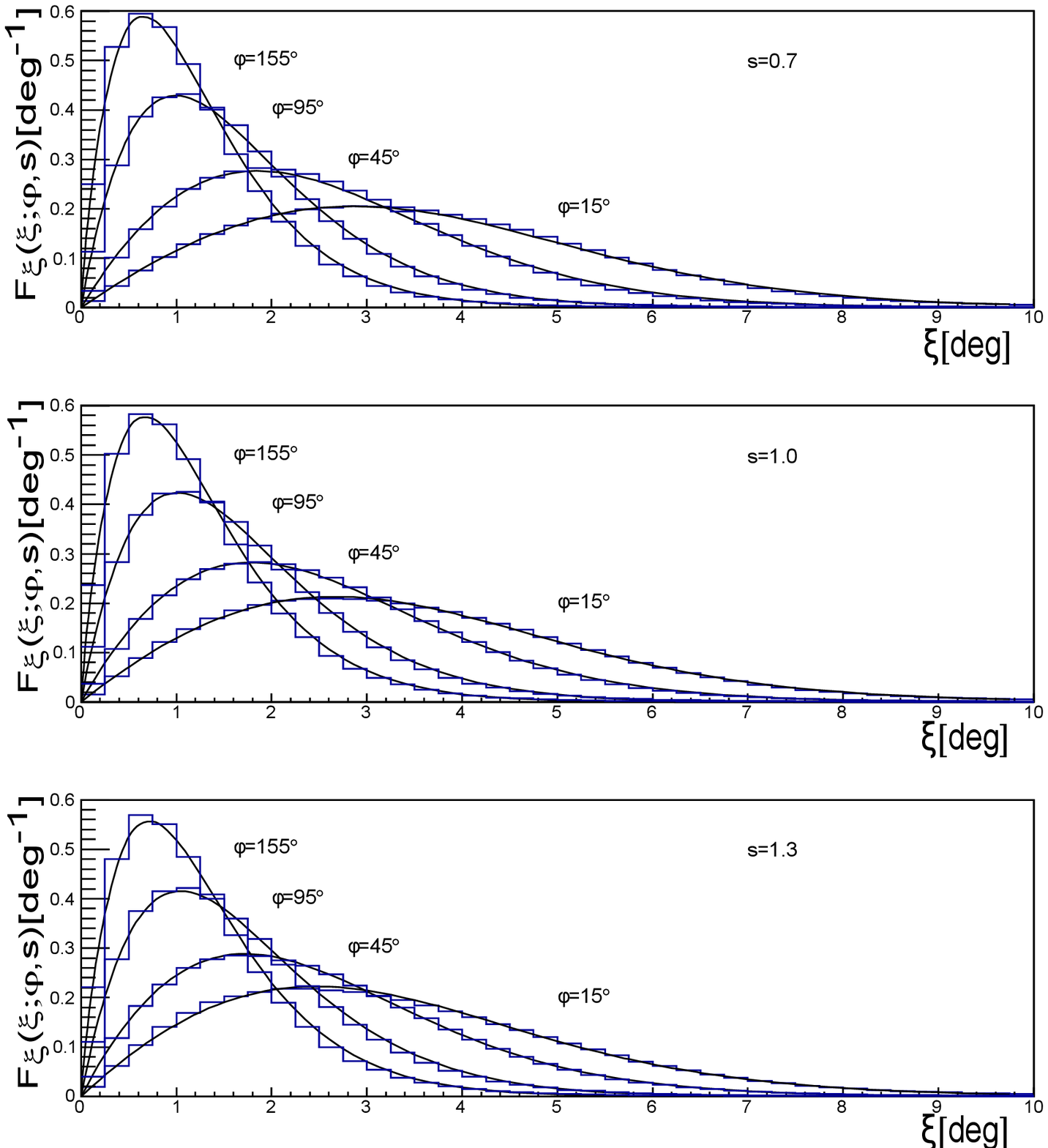}
 \caption{Comparison of the actual (histograms) and parametrized (lines) distributions $F_{\xi}(\xi;\varphi,s)$. One iron shower with $E_0=10^{17}$ eV (without thinning). }
 \label{fig9}
\end{figure}

Having established the character of the universal distributions of the azimuth angles $F_{\varphi}(\varphi;\chi, s)$ and
the distributions of the
polar angles $F_{\xi} (\xi ; \varphi , s)$, we undertake the task of finding their analytical descriptions. 
\subsection{Parametrization of the distributions of the azimuth angles $F_{\varphi}(\varphi;\chi, s)$}
The histograms in Fig. 8 present examples of the distributions
  $F_{\varphi} (\varphi ; \chi, s =1 )$ in the log-lin scale. The distributions look more or less like a fraction of a cosine function. Thus, we look for a function of the form
\begin{equation}
log\, F_{\varphi}=A+Bcos(a\,\varphi +b)\;\textnormal{,}
\end{equation}
where the parameters $A, B, a$ and $b$ have to be fitted for different values of $\chi$. 
We fit the distributions $F_{\varphi} (\varphi ; \chi, s)$ for various $\chi$ in bins $\Delta \varphi =10^{\circ}$.
The best fitting polynomial descriptions, taking into account the dependence on $s$, are given  in Appendix.\\
In Fig. 8 the histogrammed distributions  $F_{\varphi} (\varphi ; \chi, s =1 )$ refer to various values of $\chi$ for one $10^{17}$ eV iron shower. The smooth lines represent function (6) with the parameters described in Eq. A3-A4.\\

\subsection{Parametrization of the distributions of the polar angles $F_{\xi} (\xi ; \varphi , s)$}
We fit the distributions $F_{\xi} (\xi ; \varphi, s)$ for various $\varphi$ in bins $\Delta \varphi =10^{\circ}$, with the
function
\begin{equation}
F_{\xi}(\xi; \varphi, s)=\frac{C\xi}{(d+e^{\xi/2})^{\gamma}} \, \textnormal{.}
\end{equation}
 The parameters $d$ and $\gamma$ depend on $\varphi$, as it is obvious from Fig. 7. Parameter $C$ normalizes the integral $\int F_{\xi} ( \xi ; \varphi,s) d \xi $ to unity. 
The best fitting polynomial descriptions are in Appendix.\\
Fig. 9 presents an example of the distribution $F_{\xi} (\xi ; \varphi, s)$ and fitting functions for four values of azimuth angles $\varphi$ and three values of $s$. It can be seen that the description by formula (7) is quite satisfactory (see also Section 4.4).

\subsection{Estimation of the accuracy of the universal  fit}
\subsubsection{General considerations}
The analytical description of the electron distributions found in this paper, the universal fit, is not, of course, ideal. Now we aim at estimating how well it  describes the actual electron distributions in an arbitrary shower  detected by the fluorescence method as it is done e.g. at The Pierre Auger Observatory and any other similar experiment. \\
The reasons why the actual distributions and the universal fit may differ are several:
\begin{itemize}
\item  the universal fit  is meant to describe an average from several simulated proton and iron initiated showers with primary parameters somewhat arbitrarily chosen;
whereas a measured  shower has a single primary energy $E_0$, primary mass and  zenith angle arising from the distributions of these parameters in the sample registered by the experiment;
\item  the simulated distributions have been fitted with possibly simple analytical functions, what might not be good enough in all regions of the variables;
\item  there is some inaccuracy in determining a particular shower age from the simulated  shower, so that the derived electron distributions may refer to slightly different age than the chosen one;
\item  even for the same primary characteristics of the shower the electron distributions may differ in different showers a bit  above Poisson fluctuations due to some intrinsic fluctuations (see Section 4.4.3), although the exact universality would mean that they do not.
\end{itemize}
\subsubsection{Applicability to Auger showers}
First, we shall check how well our universal fit describes the electron distributions in real (simulated) showers with primary  energy $E_0=10^{17}$ eV at three ages $s=0.7$, $1$ and $1.3$.
To do it we must build a  sample playing a role  of the registered showers to be compared with our fit. We shall not  use now the same sample which served for the derivation of the universal fit; it contains smaller number of showers that we demand now, they are all vertical and do not correspond to those registered by Auger.  Treating it rigorously we should simulate  showers from the  distributions of the primary energy $E_0$  and the zenith angle registered by Auger, with some adopted primary masses. However, a randomly chosen $E_0$  might be larger than the maximum energy to fully simulate a shower without the thinning procedure which introduces rather large, unwanted, fluctuations. Moreover, our fit has been done for vertical showers and although the electron component at the same atmospheric depth should not depend on the zenith angle, some small differences may be present due to e. g. electrons from muon decay. Also the primary masses are not known with good confidence.\\
Taking all this into account we constrain ourselves to showers with ten primary energies drawn from the Gaussian distribution with $\sigma=10^{17}\times 0.14$ eV (assuming the uncertainty in the energy scale of 14$\%$ \citep{pao}) and the mean $E_0 = 10^{17}$ eV, and some arbitrarily adopted primary mass composition: half of protons and half of iron nuclei. Showers with each energy are simulated at three zenith angles $35^{\circ},45^{\circ}$ and $55^{\circ}$ (we have checked that the Auger uncertainty  of the  zenith angle, $\sim 1^{\circ}$, affects the distributions much less than that of $E_0$) so that the total number of the showers in our sample equals $n=30$. \\
For a fixed age $s$ we choose the regions of the variables $(\varphi,r/r_M,E)$
where we want to check on the accuracy of the fit of the $\theta$ distributions.
These regions are chosen so as to contain relatively many electrons.
We simulate the showers for the above primary characteristics and
obtain the distributions $\Delta N_i (\theta; \varphi,r/r_M,E, s)$ for $i=1,2,....,n$.\\
To estimate how the distributions $ \Delta N_{i} (\theta; \varphi,r/r_M,E, s)$ are described by the universal, parametrized fit
$\Delta N_{par} (\theta; \varphi,r/r_M,E, s)$ we apply the  ${\chi^2}$ test, the meaning of which in our case is the following.\\

The sample of 30 showers is believed to be drawn from the  true distribution of $\Delta N (\theta,\varphi, r/r_M,E, s)$
(for each set of the variables), with the true mean value
\begin{equation}
\Delta N_{sh}(\theta,\varphi, r/r_M,E, s) \approx \langle \Delta N (\theta,\varphi, r/r_M,E, s)\rangle  \;\textnormal{,}
\end{equation}
being close to the mean from the sample, and variance
\begin{equation}
\sigma^{2}_{sh} (\theta,\varphi, r/r_M,E, s)= \sum_{i=1}^n 
(\Delta N_i-  \langle \Delta N \rangle)^2/(n-1)  \;\textnormal{,}
\end{equation}
We want to check whether the parametrized values
$\Delta N_{par} (\theta, \varphi,r/r_M,E, s)$ can be described as  drawn from the above distribution. In other words,  we treat here $\Delta N_{par} (\theta, \varphi,r/r_M,E, s)$ in the same way as a random number $\Delta N_i (\theta, \varphi,r/r_M,E, s)$.\\
Assuming that $\Delta N (\theta, \varphi,r/r_M,E, s)$ has  Gaussian distribution  the random variable ${\chi^2}$  defined as
\begin{equation}
\chi^2=\sum(\Delta N_{par} - \Delta N_{sh} )^2/\sigma^{2}_{sh}(\theta,\varphi, r/r_M,E, s)  \;\textnormal{,}
\end{equation}
where the sum is over all $m$  cells of the variables $\theta,\varphi, r/r_M,E, s$,
has the ${\chi^2}$ distribution with $m-3$ degrees of freedom (two parameters, the mean and the variance, are determined from the sample data), provided that all  $m$ random values $\Delta N_{par} (\theta,\varphi, r/r_M,E, s)$ are independent of each other (which, of course, is not quite true). We have taken into account only cells with $\Delta N>100$.\\
The ${\chi_1^2}$ for one degree of freedom equals
\begin{equation}
 {\chi^{2}}_{1}=\frac{1}{m-3}\sum(\Delta N_{par}-\Delta N_{sh})^2/\sigma^{2}_{sh} \;\textnormal{,}
\end{equation}
We obtain that ${\chi^{2}}_{1}=0.47$, meaning that the deflections of our fit are well contained within the fluctuations between the chosen showers. So, the accuracy of the fit  looks good enough for the reconstruction of the Auger showers at $\sim10^{17}$ eV. 
However, we should keep in mind that assumption of the independence of the fit values in the neighboring cells is not valid. The independence might be between groups of several cells, being equivalent to a smaller number of degrees of freedom, resulting in turn with a larger $\chi^2$.  

\begin{figure}[!htb]
        \centering
        \includegraphics[width=4.8in]{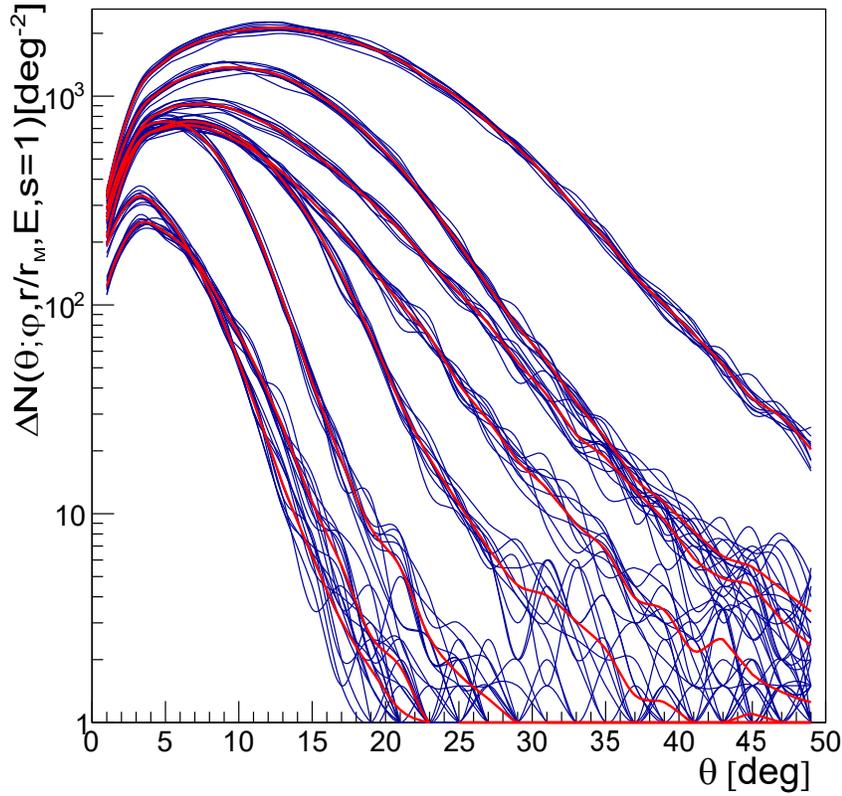}
 \caption{Number of electrons in four-dimensional cells $\Delta\,V=d\,log\,E\cdot d\,log(r/r_M)\cdot d\,\varphi\cdot d\,\theta$ at $s=1$ for a set of different values of $\varphi$, energy and lateral distances. Each of the eight groups of curves refers to 10 Fe showers with $E_0=10^{17}$ eV, thick curves being averages, for $\varphi=15^{\circ}$ and $85^{\circ}$, $E=23$ MeV and 72 MeV, $log\,r/r_M=-0.95$ and $-0.45$.}
 \label{fig10}
\end{figure}
\subsubsection{Compatibility with true distribution}
As a measure of the accuracy of our parametrization independent of any experiment we shall use a mean relative difference between the parametrization predictions and the "true" values of $\Delta N_i$, the latter obtained from shower simulations for a fixed primary energy. We obtain
\begin{equation}
\Bigg\langle \frac{|\Delta N_{par}-\Delta N_{sh}|}{\Delta N_{sh}}\Bigg\rangle = 0.059 \;\textnormal{,}
\end{equation}
where $\Delta N_{sh}$ are the previously used averaged numbers from the set of 30 showers (Section 4.4.2) and the average is taken over all cells.\\
To compare it with the minimum fluctuations of the electron multidimensional densities, i. e. those in a sample of simulated showers with the same primary parameters, we have simulated 10 vertical iron showers with  $E_0=10^{17}$ eV and calculated the mean relative dispersion of $\Delta N_i$ averaged over the variable space. Our result for all three ages is: 
\begin{equation}
\Big\langle \frac{\sigma_{sh}}{\Delta N_{sh}}\Big\rangle = 0.064 \;\textnormal{,}
\end{equation}
with values 0.073, 0.051, 0.071 for $s = 0.7$, 1, 1.3 respectively.\\
Thus, the accuracy of our parametrization is roughly the same as the level of minimal shower fluctuations $\sim 6\%$.
We have used only the cells with $\Delta N_{sh} > 100$. Cells with lower number of electrons do not have much physical meaning but add a lot to the dispersions $\sigma_{sh}$ (Fig. 10).\\
The Poisson fluctuations in the chosen cells are less than 0.1. Subtracting them in each cell from $\sigma_{sh} /\Delta N_{sh}$ in quadrature and averaging over all cells with $\Delta N_{sh}>100$ we obtain for the intrinsic relative fluctuations 
\begin{equation}
\Big\langle \frac{\sigma_{intr}}{\Delta N_{sh}}\Big\rangle = 0.028 \;\textnormal{,}
\end{equation}
with values   0.038, 0.018, 0.030 for $s=0.7$, 1, 1.3 respectively.
Since the Poisson fluctuations have been subtracted we expect that these numbers should not depend on the choice of the cell volume $\Delta V$, as long as it is small. For proton initiated showers the fluctuations should be larger. With another choice of the variable volume one can add the corresponding Poisson fluctuation obtaining the best accuracy with which fitting the real distributions makes sense.
\section{Influence of the geomagnetic field}
The shower simulations which we have done were for the geomagnetic field switched off, $B = 0$. This field switched on would change the directions of electrons and add a new variable to describe the angular distributions of electrons \citep{homola}. As a result also the lateral distributions are changed. The effect has no universal character as it depends on the local magnetic field. It can be allowed for, if needed, by deforming the universal distributions. 
We shall estimate how big this deformation is on particular levels in the atmosphere without shower simulations, by a simple analytical method, using some approximate assumptions. We shall also give a recipe how to obtain the corrected distributions from those without field $\mathbf{B}$ when the effect is small.\\
So far by electrons we have meant particles of both signs. In this section, however, "electrons" means negative-sign particles. For a deflection angle $\alpha\ll 1$ the new electron angles $\theta '$ and $\varphi '$ are 
\begin{equation}
\theta '=\theta+\Delta \theta=\theta +\cos\, \varphi\,\alpha \, \textnormal{  ,  }
\varphi '=\varphi+\Delta \varphi= \varphi -\frac{\sin\,\varphi}{\sin\, \theta}\,\alpha \quad \textnormal{,}
\end{equation}
with $\alpha>0$ for $e^{-}$ and $\alpha<0$ for $e^{+}$,
assuming that $\cos\,\theta \sim 1$ and the azimuth angle $\varphi$ is measured from the direction perpendicular to $\mathbf{B_{\perp}}$ clockwise. Thus, the new angular distribution functions $f'_{\theta}$ and $f'_{\varphi}$ are
\begin{eqnarray}
f'_{\theta}(\theta;\varphi,r/r_M,E,s)=f_{\theta}[(\theta-\cos\,\varphi\,\alpha);\varphi,r/r_M,E,s]\quad \textnormal{and}       \\
f'_{\varphi}(\varphi;r/r_M,E,s)=f_{\varphi}[\varphi+\frac{\sin\,\varphi}{\sin\, \theta}\,\alpha;r/r_M,E,s]	 \quad \textnormal{,} \nonumber
\end{eqnarray}
since $\Delta\theta'=\Delta \theta$ and $\Delta\varphi'=\Delta\varphi$.\\
A similar way is applied to find the new lateral function $f'_r$. If the new radial vector $\vec{r} ' = \vec r +\vec \rho$ then the change of the radial distance equals $\Delta r= \rho\,cos\,\varphi_S$, where $\rho>0$ for $e^{-}$ and $\rho<0$ for $e^{+}$ and $\varphi_S$ is the azimuth angle of $\mathbf{r}$ around the shower axis, measured from the direction perpendicular to $\mathbf{B_{\perp}}$ clockwise. Thus, the new function $f_r'$ depends on the azimuth $\varphi_S$. Since all electrons with energy $E$ at $r-\rho\, cos\,\varphi_S$ on a surface with area $\Delta S$ are shifted by the same distance $\rho$ perpendicularly to $\mathbf{B_{\perp}}$ we have that 
\begin{equation}
\frac{f'_{r}(r/r_M,\varphi_S;E,s)}{2\pi(r/r_M)}\Delta S'= \frac{f_{r}(r/r_M-\rho/r_M\,\cos\varphi_S;E,s)}{2\pi(r/r_M-\rho/r_M\,\cos\varphi_S)}\Delta S\, \textnormal{  , where }
\Delta S'=\Delta S \quad \textnormal{.}
\end{equation}
Thus, our recipe for allowing for the geomagnetic effect in the lateral distribution function is the following:
\begin{equation}
f'_{r}(r/r_M,\varphi_S;E,s)\simeq f_{r}(r/r_M-\rho/r_M\,\cos\varphi_S;E,s)(1+\frac{\rho\,\cos\varphi_S}{r})\, \textnormal{. }
\end{equation}
Concerning the new functions for electrons of both signs we have 
\begin{equation}
f'_{v}(v;...)=P^{-}(E;s)f^{-'}_{v}(v;...)+ P^{+}(E;s)f^{+'}_{v}(v;...)     \, \textnormal{, }
\end{equation}
where $v$ stands for variable $r$ or $r/r_M$, $\varphi$ or $\theta$ , $P^{-/+}(E;s)$ are fractions of  electrons (positrons) in their total number with energy $E$ at shower age $s$ , $f^{-'}_{v}(v;...),f^{+'}_{v}(v;...)$ are the above determined functions for electrons $(\alpha, \rho > 0)$ and positrons $(\alpha, \rho < 0)$ respectively.
At shower maximum $P^{-/+} (E;s)$ are equal $0.5$ only at $E > 300$ MeV, the fraction of positrons going down to $\sim 1/3$ at 10 MeV (see e.g. Lafebre et al. 2009). \\
 It remains to determine the values of angular and lateral deflections  $\alpha$  and  $\rho$.\\
We shall consider electrons and positrons with a given energy $E$ on a level with age $s$. However, as it has been already established (Giller et al. 2005a) the angular distributions of  electrons of both signs with fixed energy do not depend on shower age $s$. They depend solely on the electron energy $E$, so that the shower age becomes irrelevant.\\
 Hillas (1982) considered this problem by simulating small energy electromagnetic cascades developing in the magnetic field, using detailed formulae for relevant cross sections. For any electron energy $E$ he obtained an effective path length along which field $\mathbf{B}$ should act, as if electron had constant energy $E$, to obtain the correct  mean deflection angle. For $E >20$ MeV the effective path lengths were a fraction of one radiation unit $(r.\, u.)$, the largest being $0.73\,\, r.\,u.$. at $E \to \infty$. Homola et al. (2015) studied azimuthal asymmetry caused by the geomagnetic field of the electron angular distributions, integrated over the lateral spread and energy, in hadronic showers of highest energies. They obtained that the effect may be quite big in very inclined showers developing high in the atmosphere. Here we study this effect for fixed lateral distances and fixed electron energies, where the angular and lateral distributions may be distorted, as it will turn out, by a rather small amount (what, however, may be necessary to allow for).\\ 
Considering an electron along its way back in a shower one comes at some point to a photon, then again to electron (or positron), then again to electron or positron, and so on. Photons are not deflected and positron deflection decreases that accumulated by the electron. Calculating analytically the total deflection in this case, taking into account the energy losses, is not the aim of this paper (if easily possible). However, we can find the upper limit for it by considering the magnetic deflection of an electron keeping his identity all the time and losing energy on bremsstrahlung and ionisation in an average way:
\begin{equation}
-d\,E/d\,t=E+\epsilon \, \textnormal{, }
\end{equation}
$t$ being the path length in radiation units and $\epsilon=82$ MeV is the critical energy of the air. The solution is
\begin{equation}
E(l;E_{0})=(E_{0}+\epsilon)\,\exp\,(-l/l_0)-\epsilon\, \textnormal{, }
\end{equation}
with $l/l_0=t$, where $l_0$ is the radiation unit in meters and $E_0$ - the initial electron energy. \\
The magnetic deflection angle $\Delta\varphi$ of a relativistic electron with energy $E$ along path length $\Delta l \ll R(E)$, with $R(E)$ being the curvature radius,  equals 
\begin{equation}
\Delta \varphi=0.03\frac{B_{\perp}(\textrm{G})}{E(\textrm{MeV})}\Delta l(\textrm{m})=1.72^{\circ}\frac{B_{\perp}(\textrm{G})}{E(\textrm{MeV})}\Delta l(\textrm{m})\, \textnormal{. }
\end{equation}
Thus, on the way from $E_0$ to $E$, in a constant magnetic field, the electron is deflected by
\begin{equation}
 \varphi= B_{\perp}\int\frac{dl}{E'(l;E_0)}=B_{\perp}\frac{l_0}{\epsilon}\Big(\ln\frac{E+\epsilon}{E}-\ln\frac{E_0+\epsilon}{E_0}\Big)\, \textnormal{. }
\end{equation}
For $E_0\to \infty$  
\begin{equation}
\varphi\to B_{\perp} \frac{l_0}{\epsilon}\ln\frac{E+\epsilon}{E}=\varphi(\epsilon, l_0)\ln\frac{E+\epsilon}{E}=5.84^{\circ}B_{\perp}(\textrm{G})\ln\frac{E+\epsilon}{E}\, \textnormal{, }
\end{equation}
 at sea level,
where $\varphi(\epsilon,l_0)$ denotes the deflection angle of an electron with constant critical energy $\epsilon$ of the air along one radiation unit $l_0$, adopted as $36.2\, \textrm{g cm}^{-2}$.\\ 
Hillas determined the effective path length $x_M (E)$ in low energy electromagnetic cascades but due to the universality of electron distributions the same path lengths should also apply to the high energy hadronic showers and we shall use them below.   \\
The effective path length $x_M (E)$ as defined by Hillas, equals in our case 
\begin{equation}
x_M(E)=E\int\frac{dl}{E'(l;E_0)}=l_0\frac{E}{\epsilon}\,   \ln\frac{E+\epsilon}{E}{\underset{E\to\infty}{\rightarrow}}l_0\, \textnormal{. }
\end{equation}
As should be expected it is larger than that obtained by cascade simulations of Hillas where $x_M(E)\underset{\scriptstyle E \to \infty}{\to} 0.73\,l_0 $, where the negatively charged electron progenitors are also photons and positrons not increasing its deflection.\\
The geomagnetic field at earth is a fraction of 1 G. At the Auger site $B = 0.246$ G. Due to its rather large declination angle of $-35^{\circ}$, the average $B_{\perp}$ for showers arriving isotropically with zenith angles smaller than $60^{\circ}$  is not much smaller, $B_{\perp}\approx 0.9\,B\approx 0.22$ G. Thus, our estimation of the deflection upper limit for e. g. $E = 50$ MeV would be $1.6^{\circ}$ (adopting air density at Auger $1\cdot 10^{-3} \textrm{g cm}^{-3}$). The actual deflection, however, is smaller and according to Hillas the effective path length $x_M (E=50\,  \textrm{MeV}) = 12.35\, \textrm{g cm}^{-2}$, what is factor of $1.73$ smaller than the upper limit (Eq. 25). Thus, the actual average deflection is $\alpha \sim 1^{\circ}$ at Auger level. At higher atmospheric levels $\alpha=1^{\circ}\exp(h_A/8.5\,\textrm{km})$ where $h_A$ is the height above Auger site ($\sim 1.4\, \textrm{km}\, \textrm{a.s.l.}$). \\
 At the same time the electron would acquire by Coulomb scattering an angle $\theta$ with respect to the shower axis which we can estimate by a similar way as above. Adopting the average increase rate of $\langle \theta^2 \rangle$ as
\begin{equation}
\frac{\langle d\,\theta^2 \rangle}{dt}=\Big(\frac{21}{E(\textrm{MeV})}\Big)^2\, \textnormal{, }
\end{equation}
and the energy losses from Eq. 20 we obtain for $E_0\to\infty $
\begin{equation}
\sqrt{\langle \theta^2 \rangle}=\frac{21\,\textrm{MeV}}{\epsilon}\sqrt{\frac{\epsilon}{E}-\ln\Big(\frac{E+\epsilon}{E}\Big)}\, \textnormal{, }
\end{equation}
what gives $\sqrt{\langle \theta^2 \rangle}\sim 12^{\circ}$ for $E = 50$ MeV. Thus, in comparison with the calculated above $1.6^{\circ}$ magnetic shift of such electron its Coulomb scattering looks quite big. It is worth to note, however, that for actual electrons in a shower, both angles will be smaller, but their ratio, Coulomb to magnetic, will even increase since the former decreases due to the parent photons and the latter - due to photons and positrons.\\ 
From the above we may draw a conclusion that the effect is rather small for typical electron parameters, at least at heights $h_A < 20$ km where $\alpha \lesssim 10^{\circ}$, so that we can find the relative change of the angular distribution as follows:
\begin{equation}
\frac{\Delta(f_{\varphi}f_{\theta})}{f_{\varphi}f_{\theta}}= \frac{\Delta f_{\varphi}}{f_{\varphi}}+\frac{\Delta f_{\theta}}{f_{\theta}}=-\frac{1}{f_{\varphi}}\frac{\partial f_{\varphi}}{\partial \varphi}\frac{\sin\,\varphi}{\sin\,\theta}\alpha+
\frac{1}{f_{\theta}}\frac{\partial f_{\theta}}{\partial \theta}\cos\,\varphi\,\alpha\, \textnormal{, }
\end{equation}
 where we have assumed that the number of electrons $\Delta N(r/r_M;E,s)$ stays unchanged. The relative change of the angular distribution function depends strongly on the choice of angles where the derivatives are taken (see Figs. 1-4). For some representative values, $s=1,\, E=50\,\textrm{MeV},\, log(r/r_M) \sim -0.7 $, and regions of angles where the derivatives are more or less constant, we get 
\begin{equation}
\frac{\Delta(f_{\varphi}f_{\theta})}{f_{\varphi}f_{\theta}}= (0.025\frac{\sin\,\varphi}{\sin\,\theta}-0.085\,\cos\,\varphi)\,\alpha\, \textnormal{, }
\end{equation}
where $\alpha$ is in degrees. \\
 Using the effective path length for an electron with energy $E$ we have that the lateral shift $\rho$ equals
\begin{equation}
\rho=\frac{1}{2}\frac{l^2_{eff}(E)}{R(E)}\, \textnormal{, }
\end{equation}
$R(E)$ being the electron curvature radius.
Thus, an electron with $E=50$ MeV and $l_{eff}=12.35\,\textrm{g cm}^{-2}$ \citep{hillas} will be deflected on average by $\rho \sim 1$ m at the Auger level. Using the shape of the lateral distribution function $f_r$ (Giller et al. 2015) we obtain for $s=1$ and $log(r/r_M)=-0.7$ 
\begin{equation}
\frac{\Delta f_r}{f_r}= 0.003\,\cos\,\varphi_S\, \rho(\textrm{m})\, \textnormal{. }
\end{equation}
The total relative change of the number of electrons for our example, adopting $\theta =10^{\circ}$, equals
\begin{equation}
\frac{\Delta(f_rf_{\varphi}f_{\theta})}{f_r f_{\varphi}f_{\theta}}=  0.003\cos \varphi_S \,\rho(\textrm{m})+
(0.144\sin \varphi - 0.085\cos \varphi)\alpha(^{\circ})\, \textnormal{, }
\end{equation}
where for the Auger site $\rho=\pm 1\,\textrm{m}\, \exp(h_A/8.5\,\textrm{km})$, $\alpha=\pm 1^{\circ}\,\exp(h_A/8.5\,\textrm{km})$ and sign $+$ is for $e^{-}$. For electrons of both signs we have
\begin{eqnarray}
\frac{\Delta(f_rf_{\varphi}f_{\theta})}{f_r f_{\varphi}f_{\theta}}= P^{-}(E;s)[0.003\cos \varphi_S\, \rho(\textrm{m})+
(0.144\sin \varphi - 0.085\cos \varphi)\alpha(^{\circ})]+ \nonumber \\
P^{+}(E;s)[-0.003\cos \varphi_S\, \rho(\textrm{m})-
(0.144\sin \varphi - 0.085\cos \varphi)\alpha(^{\circ})]\, \textnormal{, }
\end{eqnarray}
The two angles $\varphi_S$ and $\varphi$ are, however, correlated. The average electron direction vector at a given distance $\mathbf{r}$ lies in the plane containing $\mathbf{r}$ and the shower axis, so that $< \varphi > = \varphi_s$.
 Adopting both angles equal we obtain for the maximum value of the total relative change 
\begin{equation}
\Big| \frac{\Delta f}{f}\Big|_{max} \approx 0.033\exp (h_A/8.5\, \textrm{km})\frac{B_\perp}{0.22\,\textrm{G}}\, \textnormal{, }
\end{equation}
where $f=f(\theta=10^{\circ}, \varphi<40^{\circ}, r/r_M=10^{-0.7}, E=50\,\textrm{MeV},s=1)$, $P^{-}(50\,\textrm{MeV}; s=1)=0.6$ and $P^{+}(50\,\textrm{MeV}; s=1)=0.4$. 
A more representative quantity is, however, the average modulus of the r.h.s. of Eq. 33 when the numerical coefficient equals $0.033\cdot 2/\pi= 0.021 $. Thus, for our example the relative change of the total distribution function is below $\sim 20\%$ up to heights $h_A \lesssim 20$ km.\\
The geomagnetic correction depends on the value of  
\begin{equation}
\exp (h_A/8.5\, \textrm{km})\frac{B_\perp}{0.22\,\textrm{G}}\propto\frac{B_\perp}{\rho_{air}(h)}=a \, \textnormal{, }
\end{equation}
where $\rho_{air}(h)$ is the air density at height $h$ of the shower level. The parameter $a$ derived here is just the same as that used by Homola et al (2015) for a parametrization of their simulation results. They calculate the asymmetry of the angular distribution of Cherenkov photons (being almost the same as that of electrons) integrated over lateral distance and electron energy. E.g. for a $10^{19}$ eV shower maximum at height $8-10$ km, with $B_{\perp} = 0.2$ G they obtain 
$\Big| \frac{\Delta f}{f}\Big|_{max} \approx 0.2$,
whereas for our example parameters (formula (34) allowing for slightly different $\theta$) we get $\sim 0.1$. 
Our result, however, can be treated as a lower limit since we have chosen the variable regions where the distribution functions have the second derivative almost zero, so that if the numbers of electrons and positrons were equal the effect would be zero. Since the result of Homola et al. refers to an average effect for all electrons emitting Cherenkov light ($E >30$ MeV at that height) both results are compatible.
\section{Method of shower reconstruction}
The universality of the electron distributions in showers opens a way to a new method for reconstructing shower development in the atmosphere, as it is achieved with the fluorescence technique \citep{fluo}. At the Pierre Auger Observatory the Fluorescence Detector measurements consist in registering shower light images within small time bins by (at least one) camera with 440 PMT pixels with $1.5^{\circ}$ field of view each. The shower reconstruction means determining the shower profile i.e. the number of charged particles, $N(X)$, as a function of the atmospheric depth $X$ from the measured time series of the light images. Since the shape of $N(X)$ changes little from shower to shower, having the form proposed by Gaisser and Hillas \citep{GHill}, it is $N_{max}$ and $X_{max}$ that determine it to a good approximation. Adopting their values and using the universal description of the electron states at any shower age allows one to predict exactly the light fluxes (both fluorescence and Cherenkov) emitted by shower electrons at consequent time intervals. We would like to stress that to do it the full function $f(\theta,\varphi, r/r_M, E;s)$ is needed for showers with measurable lateral size. Finding values of $N_{max}$ and $X_{max}$ fitting best the data one can determine the shower primary energy, 
and from the depth $X_{max}$ - obtain an estimate of the primary mass.\\
This method would be particularly useful for reconstructing showers with a small
detecting angle (angle between shower axis and camera symmetry axis), since
they contain a not negligible amount of the Cherenkov light, as these registered by an additional (at Auger) set of telescopes HEAT \citep{pao}. It is thanks to finding in this paper the universal function $f(\theta,\varphi, r/r_M, E;s)$ that this light can be treated accurately for the first time.
\section{Discussion and summary}
This paper is the last one from a series where gradually more variables describing electron state in a shower were taken into account. First, it was the electron energy spectra that were shown to be universal and dependent on the shower age only \citep{eSpec}. Then the lateral distributions expressed in Moli\`ere radius, were shown to be universal as well \citep{eas,nerl}. The new idea \citep{eas} was that the lateral distributions were studied for electrons with fixed energies and then they turned out to be independent even of the shower age. They could be described by a universal, one dimensional function. \\
Finally, the angular distributions, dependent on the electron energy and lateral distance, have been found in this paper for the first time. As one should expect, they are universal as well.
The universality has made the analytical description of the electron distributions meaningful and relatively easy. Particularly so, since the number of variables to describe them has been found to be smaller than five, disclosing a new sort of universality being an inner characteristic of showers (in Giller et al. (2015) it was called an extended universality). \\
Universality means here independence of the primary particle energy or mass, zenith angle or fluctuations from shower to shower. We do not mention the high energy interaction model since the independence of the primary mass implies the independence of the model. Indeed, one could imagine that the iron shower is actually a proton one with the interaction model corresponding to the iron nucleus.
This would be a dramatic change for the proton interactions, much bigger one than the differences between the existing models.
A demonstration of it may be the dependence of $X_{max}$ at some primary energy on the interaction model and primary mass. The difference of $X_{max}$ between proton and iron showers is considerably larger than that for different models and the same primary mass. \\
The universality of the angular distributions, dependent on shower age, electron energy and lateral distance has been checked in the primary energy region $10^{16}-10^{19}$ eV. However, since the lateral distributions are universal for energies up to $10^{20}$ eV, it would be hard to believe that the angular distributions are not. For energies lower than $10^{16}$ eV the numbers of electrons in the cells $\Delta logE \,\Delta log(r/r_M)\, \Delta\varphi\, \Delta\theta$ would be mostly too small to determine the density there. Nevertheless, there seems to be no reason why the universality should stop at $10^{16}$ eV, although it could be checked for averages of many showers only, not for single showers as it could be done at higher energies.\\ 
The distributions found here are normalized to unity. Thus, to find the actual number of electrons in a given cell of variables at a particular depth $X$ in the atmosphere one has to know the total number of electrons at this depth $N(X)$. Without going too much into details, this can be done by adopting the two free parameters of the Gaisser-Hillas function, $N_{max}$ and $X_{max}$ describing $N(X)$.\\
The universality of showers, consisting mainly of electrons, opens a new method for deriving $N(X)$ from the optical data of the observatories as The Pierre Auger Observatory and Telescope Array. They measure the light fluxes from a shower while it is propagating through the atmosphere. It is not only the fluorescence but also the Cherenkov light. From the comparison of the measured light intensities with those predicted the best fitting $N_{max}$ and $X_{max}$ can be found.
However, to predict well both light components the full electron distribution $f(\theta,\varphi, r/r_M, E,s)$ has to be known. It is particularly important for the HEAT addition to Auger, measuring closer, lower energy showers where the Cherenkov light plays a much bigger role than for the much distant Auger showers. The distributions found here, describing the electron angular distributions as function of their lateral distance and energy, allow one to predict exactly the Cherenkov contribution for the first time.  \\
At some cases a distortion of the angular and lateral distributions by the geomagnetic field would have to be taken into account. We have derived analytically when and how to allow for it.\\
Our parametrized distributions diverge from the simulated ones typically by $\sim 6\%$. It is roughly the same as the dispersion between showers with the same primary conditions. It is also well within the experimental uncertainties resulting from the accuracy of the determination of the primary energy as at the Pierre Auger Observatory.

\paragraph{Acknowledgments}
We are particularly grateful to the authors of the CORSIKA computer code, since without it our results would not be able to be achieved. We also thank the Pierre Auger Collaboration who showed us the importance of this work.
This work has been supported by the grant no. DEC-2013/10/M/ST9/00062 of the Polish National Science Center.

\appendix
\section{Parametrization of the universal angular distributions} 
\subsection{Distributions of the azimuth angles $F_{\varphi}(\varphi;\chi, s)$} 
Using the variable:
\begin{equation}
  \chi=E\cdot r/r_M  \;\textnormal{,}
\end{equation} 
where $E$ is in MeV,
we have reduced the distribution $f_{\varphi}(\varphi;r/r_M,E,s)$ of four variables to the distribution $F_{\varphi}(\varphi;\chi, s)$ of three variables, so that 
\begin{equation}
f_{\varphi}(\varphi;r/r_M,E,s)=F_{\varphi}(\varphi\,;\chi=E\cdot r/r_M, s)\;\textnormal{.}
\end{equation}
We fit that distributions $F_{\varphi}$ with the function
\begin{equation}
log\, F_{\varphi}=A+Bcos(a\,\varphi +b)\;\textnormal{,} 
\end{equation}
where $\varphi$ is in degrees and the best fitted parameters are 
\begin{eqnarray}
A&=&-2 \quad \textnormal{,}      \nonumber \\
B&=&(0.491-0.084\cdot s)+(0.482-0.178\cdot s)\cdot\log \chi\quad \textnormal{,}      \nonumber \\
a&=& (0.817-0.290\cdot s)+(-0.035+0.144\cdot s)\cdot\log \chi  \quad \textnormal{,} \nonumber \\
b&=&(58.44+22.92\cdot s)+(0.814-11.80\cdot s)\cdot \log \chi \;\textnormal{.} 
\end{eqnarray}
\subsection{Distributions of the polar angles $F_{\xi} (\xi ; \varphi , s)$}
Using the variable:
\begin{equation}
\xi=\frac{E^{\alpha}(1+r/r_M)}{(r/r_M)^{\beta}}\cdot \theta  \;\textnormal{,}
\end{equation}
where $E$ is in GeV and $\theta$ in degrees and parameters $\alpha$ and $\beta$ are
\begin{eqnarray}
\alpha&=&0.61+0.76\cdot 10^{-3}\cdot \varphi \quad \textnormal{and}       \\
\beta&=&0.46-2.1\cdot 10^{-3}\cdot \varphi+5.5\cdot 10^{-6}\cdot \varphi^2 \quad \textnormal{where}\ \varphi	 \quad \textnormal{is in degrees,} \nonumber
\end{eqnarray}
 we have reduced distribution $f_{\theta} (\theta ; \varphi , r / r_M , E , s )$ of five variables to the distribution $F_{\xi} (\xi ; \varphi,s )$ of only three variables, so that 
\begin{equation}
f_{\theta} (\theta ; \varphi , r / r_M , E , s )=F_{\xi} \Big(\xi=\frac{E^{\alpha} (1+r/r_M)\theta}{(r/r_M)^{\beta}}\, ; \varphi,s \Big)\cdot \frac{d\, \xi}{d\, \theta}\;\textnormal{.}
\end{equation}
We fit distributions $F_{\xi} (\xi ; \varphi,s)$ with the function
\begin{equation}
F_{\xi}(\xi; \varphi,s)=\frac{C\xi}{(d+e^{\xi/2})^{\gamma}} \, \textnormal{,}
\end{equation}
where  $C$ is the normalization constant and the best fitted parameters are:\\ 
\begin{eqnarray}
log\,d&=&d_1-d_2\cdot 10^{-3}\cdot \varphi+d_3\cdot 10^{-5}\cdot \varphi^2\ \textnormal{,} \nonumber \\
\textnormal{where } d_1&=& 1.177-0.066\cdot s-0.178\cdot s^2 \ \textnormal{,}\nonumber \\
d_2&=& 15.3 -0.82\cdot s-2.86\cdot s^2 \ \textnormal{,}\nonumber\\
d_3&=& 2.28-1.48\cdot s+0.31\cdot s^2 \ \nonumber \\
\textnormal{and }
\gamma&=&1.60+1.85\cdot10^{-2}\cdot \varphi -3.92\cdot 10^{-5}\cdot \varphi^2\ \textnormal{.}
\end{eqnarray}


\begin{thebibliography}{}
\bibitem[Aab et al. 2015]{pao} The Pierre Auger Collaboration: Aab, A. et al., 2015, NIMPA, 798, 172
\bibitem[Abraham et al. 2010a]{abr1} The Pierre Auger Collaboration: Abraham,  J.  et al., 2010a, PhLB, 685, 239
\bibitem[Abraham et al. 2010b]{abr2} The Pierre Auger Collaboration: Abraham,  J.  et al., 2010b, PhRvL, 104, 091101
\bibitem[Abraham et al. 2010c]{fluo} The Pierre Auger Collaboration: Abraham,  J.  et al., 2010c, NIMPA, 620, 227
\bibitem[Abreu et al. 2010]{abreu} The Pierre Auger Collaboration: Abreu, P. et al., 2013, JCAP, 1302, 026
\bibitem[Gaisser $\&$ Hillas 1977]{GHill} Gaisser, T.K. and Hillas, A.M., 1977, Proc. ICRC, Plovdiv, 8, 353
\bibitem[Giller et al. 2004]{eSpec} Giller, M., Wieczorek, G., Kacperczyk, A., Stojek, H. and Tkaczyk, W., 2004, JPhG, 30, 97
\bibitem[Giller et al. 2005a]{Sim} Giller, M., Kacperczyk, A., Malinowski, J., Tkaczyk, W.  and Wieczorek, G., 2005a, JPhG, 31, 947
\bibitem[Giller et al. 2005b]{eas}Giller, M., Stojek,  H. and Wieczorek, G., 2005b, IJMPA, 20, 6821
\bibitem[Giller et al. 2015]{gsw}  Giller, M., \'Smia\l{}kowski,  A. and Wieczorek, G., 2015, APh, 60, 92
\bibitem[G\'ora et al. 2005]{gora} G\'ora, D., Engel, R., Heck, D., Homola, P. et al., 2006, APh, 24, 484
\bibitem[Heck et al. 1998]{cors} Heck, D., Knapp, J., Capdevielle, J.N., Schatz, G., Thouw, T., 1998, Report FZKA 6019 
 (Forschungszentrum Karlsruhe GmbH, Karlsruhe Germany) https://web.ikp.kit.edu/corsika/ 
\bibitem[Hillas 1982]{hillas} Hillas, A. M., 1982, JPhG, 8, 1461
\bibitem[Hillas 1997]{hillas2} Hillas, A. M., 1997, NuPhS, 52B, 29
\bibitem[Homola et al. 2015]{homola} Homola, P., Engel, R. and Wilczy\'nski, H., 2015, APh, 60, 47
\bibitem[Lafebre et al. 2009]{laf}Lafebre, S., Engel, S. R., Falcke, H., H\"orandel, J. et al., 2009, APh, 31, 243
\bibitem[Nerling et al. 2006]{nerl} Nerling, F., Bl\"umer, J., Engel, R. and Risse, M., 2006, APh, 24, 421

\end{thebibliography}
\end{document}